\documentclass[
    reprint,
    amsmath,
    amssymb,
    aps,
    prl
]{revtex4-2}

\usepackage{graphicx}
\usepackage{braket}
\usepackage[hypertexnames=false]{hyperref}

\begin{document}

\title{Protecting Expressive Circuits with a Quantum Error Detection Code}

\author{Chris N. Self}
\email{christopher.self@quantinuum.com}
\affiliation{Quantinuum, Partnership House, Carlisle Place, London SW1P 1BX, United Kingdom}
 
\author{Marcello Benedetti}
\email{marcello.benedetti@quantinuum.com}
\affiliation{Quantinuum, Partnership House, Carlisle Place, London SW1P 1BX, United Kingdom}

\author{David Amaro}
\email{david.amaro@quantinuum.com}
\affiliation{Quantinuum, Partnership House, Carlisle Place, London SW1P 1BX, United Kingdom}

\date{July 26, 2024}

\begin{abstract}
A successful quantum error correction protocol would allow quantum computers to run algorithms without suffering from the effects of noise. However, fully fault-tolerant quantum error correction is too resource intensive for existing quantum computers. In this context we develop a quantum error detection code for implementations on existing trapped-ion computers. By encoding $k$ logical qubits into $k+2$ physical qubits, this code presents fault-tolerant state initialisation and syndrome measurement circuits that can detect any single-qubit error. It provides a universal set of local and global logical rotations that have physical support on only two qubits. A high-fidelity -- though non fault-tolerant -- compilation of this universal gate set is possible thanks to the two-qubit physical rotations present in trapped-ion computers with all-to-all connectivity. Given the particular structure of the logical operators, we nickname it the Iceberg code. We demonstrate the protection of circuits of 8 logical qubits with up to 256 layers, saturate the logical quantum volume of $2^8$, and show the positive effect of increasing the frequency of syndrome measurements within the circuit. These results illustrate the practical usefulness of the Iceberg code to protect expressive circuits on existing trapped-ion quantum computers.
\end{abstract}

\maketitle

Quantum error correction (QEC)~\cite{Gottesman_2009, Lidar_2013} encodes quantum information redundantly into a large Hilbert space where it can be protected from noise. Once achieved, large-scale fault-tolerant QEC is expected to unlock the potential of quantum algorithms in providing computational speedups on hard problems such as simulating quantum systems~\cite{Orzel_2017}, solving large systems of equations~\cite{HHL_2009}, or factoring large numbers~\cite{Shor_1997}. Unfortunately, the requirements to get there are daunting~\cite{Gidney_2021,Allcock_2022}. Although impressive progress has been made~\cite{Chen_2021,Ryan-Anderson_2021, SurfaceGoogle_2022,Postler_2022,Ryan_2022}, existing quantum computers can only implement QEC protocols on very few logical qubits and simple states. 

Studying protocols that fit within the limitations of current quantum computers to achieve early fault-tolerant quantum computation (EFTQC) is a promising avenue~\cite{Suzuki_2022}. In this regime, a useful quantum advantage is still expected for tasks such as the simulation of many-body quantum systems involving substantial amounts of entanglement~\cite{Cirac_2012} or the sampling of classically intractable distributions~\cite{Harrow_2017, Lund_2017}. As the technology advances, a steady increase in the complexity and quality of protocols should be expected before the achievement of full large-scale fault-tolerance.

A promising starting point for EFTQC is quantum error detection (QED)~\cite{Knill_2004a,Knill_2004b,Urbanek_2020}. In this approach, syndrome measurements are executed at predetermined times and the computation is discarded upon detection of an error. This sidesteps decoding algorithms altogether at the cost of an overhead in the number of times that the quantum circuit needs to be repeated. QED does not require information about the location of the error, which in turns leads to simpler syndrome measurement circuits.

\begin{figure*}[ht]
    \centering
    \includegraphics[width=\textwidth]{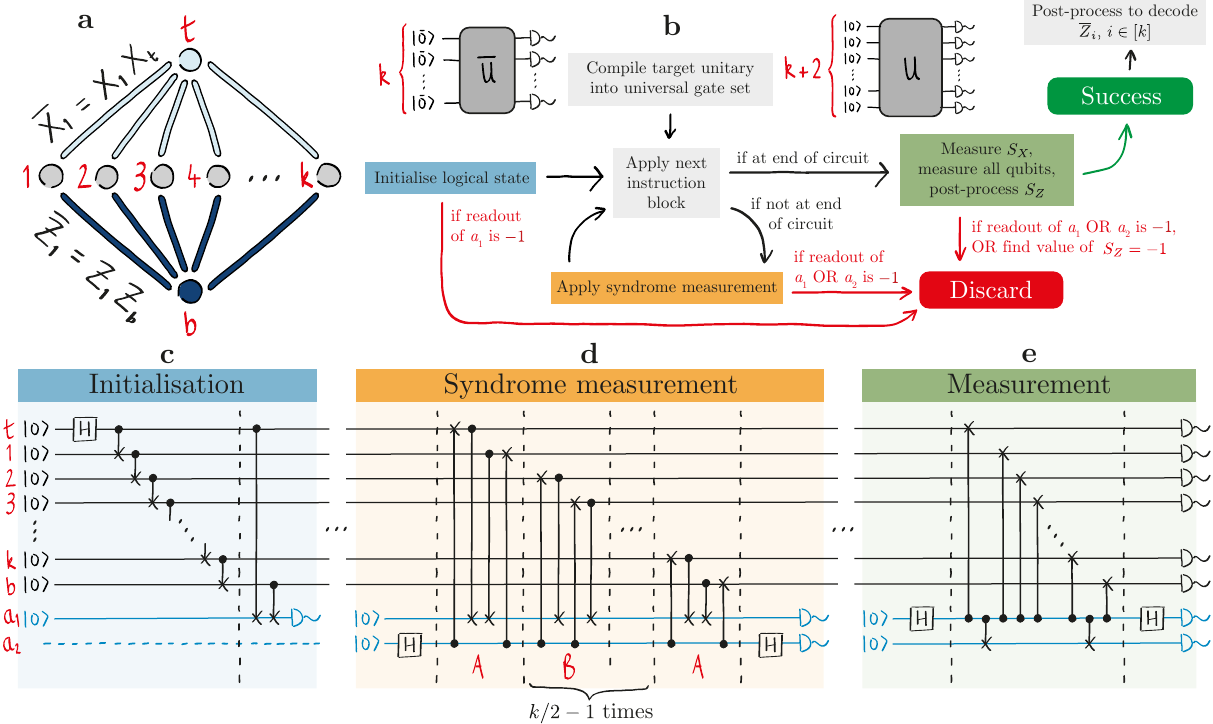}
    \caption{
    \textbf{Logical operators and fault-tolerant operations of the Iceberg code.}
    \textbf{a}~The $k$ logical single-qubit operators are defined on the $n=k+2$ physical qubits $1,\ldots,k$, $t$ and $b$. Pauli-$X$ logical operators act on $t$ and a numbered qubit, while Pauli-$Z$ logical operators act on $b$ and a numbered qubit. The two stabiliser operators $S_X$ and $S_Z$ consist on the product of $X$ and $Z$ operators on all physical qubits, respectively.
    \textbf{b}~Flowchart showing the compilation of a logical unitary $\overline{U}$ using the Iceberg code.
    \textbf{c}~Fault-tolerant initialisation of the $\ket{\overline{0}}$ logical state on the $k$ logical qubits using a single ancilla $a_1$. 
    \textbf{d}~Fault-tolerant syndrome measurement of both the $S_X$ and $S_Z$ stabiliser using two ancillas $a_2$ and $a_1$, respectively. The circuit has an $ABBB \cdots BA$ structure where $A$ and $B$ denote different orderings of the CNOTs. 
    \textbf{e}~Fault-tolerant measurement that first measures $S_X$ and then destructively measures all qubits to evaluate $S_Z$ and all logical $Z$ operators in a post-processing step.
    If at any point in the circuit the readout of $a_1$, $a_2$, or the post-processed value of $S_Z$ is $-1$ the circuit is discarded and restarted.
    }
    \label{fig:ft-circs}
\end{figure*}

Here we contribute to EFTQC with the development of the $[[k+2,k,2]]$ QED code, tailoring it towards implementations on trapped-ion quantum computers~\cite{Schindler_2013,Bruzewicz_2019}.
An even number $k$ of logical qubits are encoded into $k+2$ physical qubits with code distance 2. Such an encoding is optimal in the sense that it uses the minimum number of physical qubits~\cite{Grassl_2004}.
Another two ancillas are used for syndrome measurement, making a total overhead of only four qubits. 
We nickname it the Iceberg code because of the particular structure of the logical operators, illustrated in Fig.~\ref{fig:ft-circs}(a). Some fault-tolerant gates are known for this code~\cite{Chao2018, Gidney2023}. However, their physical implementation remains challenging and multiple applications are required to compile basic logical gates such as single-qubit arbitrary rotations.

In contrast, we note that the Iceberg code has a universal set of logical rotations that allows each rotation to be compiled directly -- though non fault-tolerantly -- into physical gates native to trapped-ion quantum computers:
one two-qubit M\o{}lmer-S\o{}rensen (MS) gate~\cite{MSgate_1999} and up to four single-qubit Clifford gates. Importantly, apart from logical gates acting on one and two logical qubits, the set includes global logical gates that act simultaneously on $k$ or $k-1$ logical qubits. Hence, the Iceberg code can compile high-fidelity global logical gates using only one two-qubit physical gate. In the spirit of EFTQC, we sacrifice the fault-tolerance of the universal gate set to obtain a simple, fast, and high-fidelity compilation of expressive quantum circuits.

To test the performance of the code, we use all 12 qubits of the Quantinuum H1-2 trapped-ion quantum computer~\cite{quantinuum-devices,Pino_2021,Ryan_2022} to implement much deeper and more expressive circuits than those of previous QED demonstrations we are aware of~\cite{Linke_2017}. Using mirror circuits~\cite{Hines_2022,proctor2021scalable} we show that the Iceberg code protects parameterised quantum circuits from noise. We then saturate the maximum logical Quantum Volume (QV) of $2^8$ that can be achieved with 12 physical qubits using the Iceberg code. Finally, we empirically demonstrate that increasing the number of syndrome measurements has a positive effect. From these results we conclude that the Iceberg code offers protection against noise to expressive quantum circuits on existing quantum computers. This shows that tailoring codes to the underlying quantum computer and algorithm of interest is a promising strategy towards EFTQC.

\section{Code definitions}

The Iceberg code is a stabilizer code that encodes $k$ logical qubits $[k] = \{1,2,\ldots,k\}$ into $n=k+2$ physical qubits $[n] = [k]\cup\{t, b\}$. There are two commuting stabilisers, both acting on all physical qubits: a Pauli-$X$ operator $ S_X=\bigotimes_{i\in[n]}X_i$, and a Pauli-$Z$ operator $S_Z=\bigotimes_{i\in[n]}Z_i$. The code space is defined as the joint $+1$ subspace of both stabilisers, so a readout $-1$ of either of them indicates the presence of an error.

Single-qubit logical operators are defined as the two-qubit physical operators $\overline{X}_i=X_iX_t$ and $\overline{Z}_i=Z_iZ_b$ for every $i\in[k]$ as illustrated in Fig.~\ref{fig:ft-circs}(a), where we have used bar notation to indicate logical operators.
As required, they pairwise anti-commute $\{\overline{X}_i,\overline{Z}_i \}=0$, commute otherwise $[\overline{X}_i,\overline{Z}_j]=[\overline{X}_i,\overline{X}_j]=[\overline{Z}_i,\overline{Z}_j]=0$, and commute with the stabilisers.

Initialisation, syndrome measurement, and final measurement circuits are fault-tolerant circuits based on the flagged circuits in Ref.~\cite{Chao_2008}, though we developed the particular $ABBB \cdots BA$ structure of the syndrome measurement circuit depicted in Fig.~\ref{fig:ft-circs}(d). The circuits require only two additional ancilla qubits, which can be reset and reused. 
Notably when extracting both $S_X$ and $S_Z$ the circuit instructions are carefully arranged so that the ancillas act as flag qubits to each other, allowing the detection of ancillary errors that would otherwise propagate to the rest of qubits. Importantly, the order of the controlled NOT (CNOT) gates in these circuits must be respected to preserve the fault-tolerance. 

In this work we follow the traditional definition of fault-tolerance described in Ref.~\cite{Gottesman_2009}. A fault-tolerant circuit is a circuit where there is no single component whose failure produces an undetectable logical error. Under this definition, the state initialisation, syndrome measurement, and final measurement circuits of the Iceberg code are fault-tolerant. We verify this by exhaustively testing every two-qubit Pauli error placed after each of the CNOT gates, every single-qubit Pauli error placed after the Hadamard gates, and single-qubit Pauli-$X$ errors placed after each state initialisation or before a qubit measurement. More details are given in the Supplementary Information.

\begin{figure}
    \centering
    \includegraphics[width=1\columnwidth]{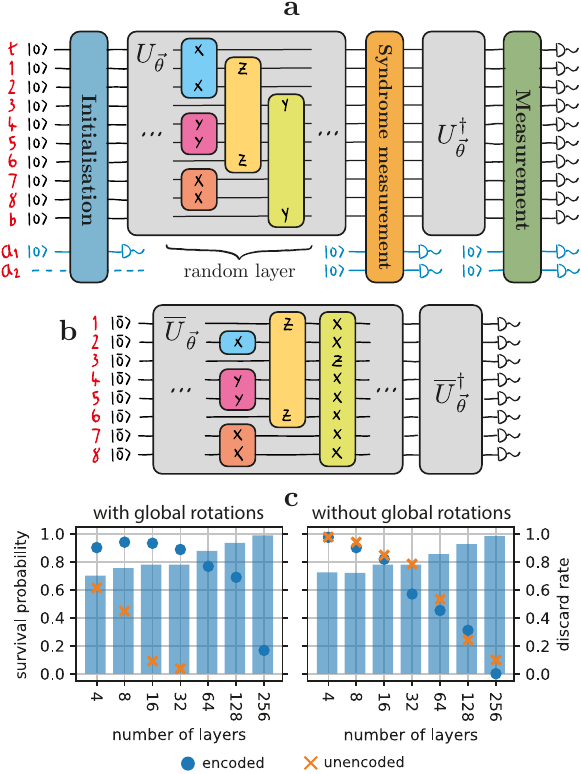}
    \caption{
    \textbf{Comparing the performance of a random mirror circuit encoded with the Iceberg code against the unencoded circuit.}
    \textbf{a}~Physical compilation of a mirror circuit on 8 logical qubits with the Iceberg code. The circuit implements a parameterised physical unitary and its gate-by-gate inverse. The unitary has a layered structure where in each layer physical qubits $[n]=\{1,\ldots,k\}\cup\{t,b\}$ are randomly paired and a random physical rotation $\exp(-i\theta \sigma_i\sigma_j/2)$ is applied on each pair $i, j\in[n]$ for $\theta \in (0,2\pi)$ and $\sigma \in \{X, Y, Z\}$. We illustrate this by showing one layer, where each physical rotation is labelled with its rotation basis.
    \textbf{b}~Corresponding logical mirror circuit. Colour is used to indicate the correspondence between each two-qubit physical rotation in (a) and logical rotation $\exp(-i\theta \overline{P}_{\sigma i j}/2)$ with logical operator $\overline{P}_{\sigma i j} = \sigma_i \sigma_j$. Note that the last logical operator $\overline{P}_{Y 3 b}$ has physical support on only two physical qubits but acts globally on all logical qubits.
    \textbf{c}~Survival probability (symbols) and discard rate (bars) plotted vs number of layers. Number of layers includes the layers in the unitary and its inverse. The survival probability is the probability of measuring the initial logical quantum state $\ket{\overline{0}}$ on all logical qubits --as expected in the absence of noise. We compare the encoded circuit, where the logical unitary $\overline{U}_{\vec{\theta}}$ is compiled with the Iceberg code as depicted in (a), and the unencoded circuit, where it is compiled into phase gadgets as described in the Supplementary Information. We differentiate logical circuits with and without global logical gates.
    }
    \label{fig:mirrortest8}
\end{figure}

Due to the particular structure of the Iceberg code, every two-qubit logical operator has support on two physical qubits: $\overline{\sigma}_i\overline{\sigma}_j = \sigma_i\sigma_j$ for every pair $i,j\in[k]$ and every $\sigma\in\{X,Y,Z\}$. Conversely, every two-qubit physical operator of the form $\sigma_i\sigma_j$ represents a different logical operator $\overline{P}_{\sigma{i}{j}}$ for every $i,j\in[n]$.
Notably, this includes global logical operators acting on $k$ or $k-1$ logical qubits in addition to the single- and two-qubit logical operators described previously. For example, $\overline{P}_{X t b}=\bigotimes_{i\in[k]}\overline{X}_i=X_tX_b$ is the tensor product of all single-qubit $X$ operators. Other examples are provided in Figs.~\ref{fig:mirrortest8}(a-b) while the Supplementary Information includes a comprehensive description of all logical operators of the form $\overline{P}_{\sigma{i}{j}}$.

We define the universal gate set of the Iceberg code as the set of logical rotations of the form $\exp(-i \theta \overline{P}_{\sigma i j}/2)$ for every angle $\theta \in (0, 2\pi)$. This universal gate set demands the all-to-all connectivity present in trapped-ion quantum computers. On the trapped-ion Quantinuum H1-2 device each logical rotation in the set can be compiled non fault-tolerantly using only one MS gate of the form $\text{MS}_{ij}(\theta) = \exp(-i\theta Z_i Z_j/2)$ and up to four single-qubit Clifford gates. Our compilation of logical operators is not fault-tolerant because two-qubit Pauli errors of the form $X_iX_j$, $Y_iY_j$, or $Z_iZ_j$ after a $\text{MS}_{ij}(\theta)$ gate produce an undetectable logical error for any pair $i,j\in[n]$. Nevertheless, in the absence of a QEC/QED protocol, every two-qubit Pauli error (not just this reduced set) after a MS gate produces an undetectable logical error. Since two-qubit logical rotations have a compilation that uses only one MS gate in both the unencoded circuit and the circuit encoded with the Iceberg code, it is expected that the code provides a higher-fidelity implementation. Hence, this reduction in the size of the problematic set of errors provides an intuitive understanding of the protection that the Iceberg code offers despite the non fault-tolerance of the logical gates. 

Consider a general circuit where $k$ logical qubits are prepared in their $\ket{\overline{0}}$ computational state, acted on by a unitary $\overline{U}$, and then measured in the computational basis. The practical steps to protect a quantum circuit with the Iceberg code are illustrated in Fig.~\ref{fig:ft-circs}(b): compile the unitary into the universal gate set, divide the resulting logical circuit into blocks of instructions, implement the initialisation circuit of Fig.~\ref{fig:ft-circs}(c), apply each block of instructions followed by the syndrome measurement depicted in Fig.~\ref{fig:ft-circs}(d), and finally measure all physical qubits. The last syndrome measurement can be combined with the measurement of all qubits as indicated in Fig.~\ref{fig:ft-circs}(e). Now, the output must be post-processed to evaluate the stabiliser $S_Z$ and the logical operators $\overline{Z}_i$. If the readout of every ancilla and the post-processed value of $S_Z$ is $+1$, consider the value of the logical operators as valid. Otherwise discard and restart the circuit. Intermediate measurements of logical qubits can be also included using the protocol described in Ref.~\cite{Chao_2008}.

\section{Experimental results}
To demonstrate the protection that the Iceberg code offers to expressive quantum circuits we conduct two experiments with 8 logical qubits encoded into the 12-qubit Quantinuum H1-2 trapped-ion quantum computer~\cite{quantinuum-devices}.

\subsection{Mirror circuits}
In the first experiment we study parameterised mirror circuits drawn from a highly expressive family of random circuits. Mirror circuits are employed in sophisticated randomized benchmarking techniques~\cite{proctor2021scalable, proctor2022measuring}. They are typically composed of Clifford gates allowing them to be efficiently analysed. Here, we preserve the mirror circuit structure, but compose it of more general arbitrary-angle parameterized gates. Figures~\ref{fig:mirrortest8}(a-b) explain our construction of these circuits and provide examples of the correspondence between physical and logical operators.

We vary the number of layers from 4 to 256 and compare the survival probability of the unencoded circuits and the circuits encoded with the Iceberg code. That is, the probability of measuring the initial logical state after implementing a logical unitary and its inverse. In the absence of noise the survival probability is 1 and it decreases with the number of errors in the circuit. Two kind of logical circuits are analysed: those that contain global logical gates and those that do not. This analysis separates the effect of the compression of global gates from the effect of encoding and error detection. We run experiments on two randomly chosen circuits; one with global rotations and one without. In the Supplementary Information we give numerical evidence that these are typical instances.

With global logical gates we are effectively implementing a more favourable scenario for the Iceberg code. Indeed the compilation of global logical gates with the code is performed by a single two-qubit MS gate, while two staircases of CNOT gates involving all qubits are necessary in the unencoded compilation. These compilations are described in more detail in the Supplementary Information. In Fig.~\ref{fig:mirrortest8}(c), left panel, we see that the Iceberg code considerably improves the survival probability of circuits with global rotations.

Without global rotations, the unencoded circuit is much simpler than the encoded circuit and does not have the overhead of the initialisation, syndrome-measurement and final measurement circuitry. For example, at 128 layers, the unencoded circuit has depth 331, including 372 two-qubit gates, whereas the encoded circuit has depth 549 with 657 two-qubit gates when compiled into native hardware gates. In Fig.~\ref{fig:mirrortest8}(c), right panel, we see that the Iceberg code gives a satisfactory survival probability even for circuits without global rotations. It is important to emphasise that this is a worst case scenario for the Iceberg code, yet we observe good results thanks to the encoding and error detection.

\begin{figure}
    \centering
    \includegraphics[width=1\columnwidth]{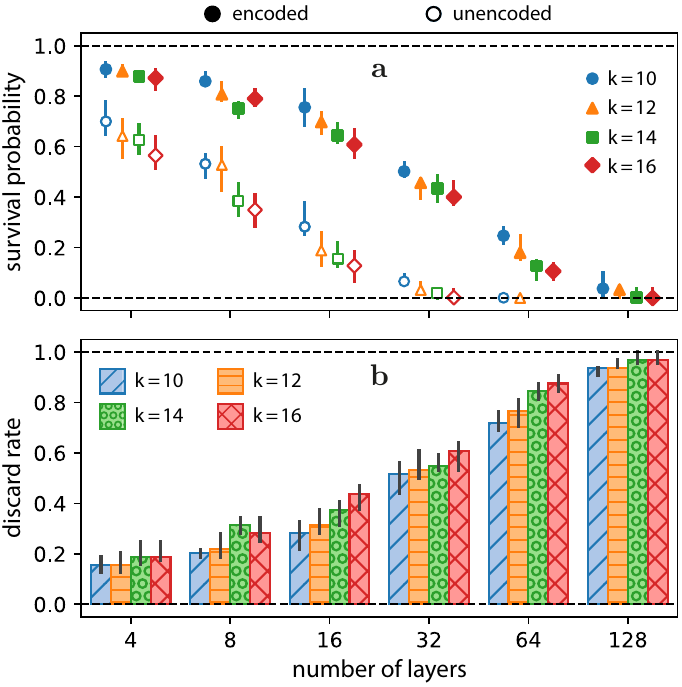}
    \caption{
    \textbf{Numerical simulations of the performance of random mirror circuits for more logical qubits.}
    We simulate 32 instances of the random circuits described in Fig.~\ref{fig:mirrortest8}, including global rotations.
    \textbf{a}~Median survival probability is plotted against number of layers for (filled symbols) encoded circuits and (hollow symbols) unencoded circuits for different numbers of logical qubits. 
    \textbf{b}~Median discard rates for the encoded circuits in each case. Mid-circuit rounds of syndrome measurement are applied every 16 layers.
    Error bars in (a) and (b) show the 99\% confidence interval on the median, obtained from bootstrap resampling.
    }
    \label{fig:scaling}
\end{figure}

To investigate the effect of adding layers and logical qubits on performance we carry out numerical simulations. We use circuits with global rotations and a simplified error model based on the specifics of H1-2~\cite{Ryan_2022}. Full details are given in the Supplementary Information.
Figure~\ref{fig:scaling}(a) shows that for a fixed survival probability the encoded circuit (filled symbols) can have approximately four times more layers than the unencoded circuit (hollow symbols). For small number of layers, adding qubits leads to a larger performance gap between encoded and unencoded circuits. The price to pay for using the Iceberg code is given by the discard rate. Figure~\ref{fig:scaling}(b) shows an increase in discard rate as we add qubits and layers. This can be compensated by increasing the number of circuit repetitions. In summary this analysis shows the usefulness of the Iceberg code for a large family of highly non-trivial circuits, and for the current generation of trapped-ion quantum computers.

\subsection{Logical Quantum Volume}

\begin{figure}
    \centering
    \includegraphics[width=1\columnwidth]{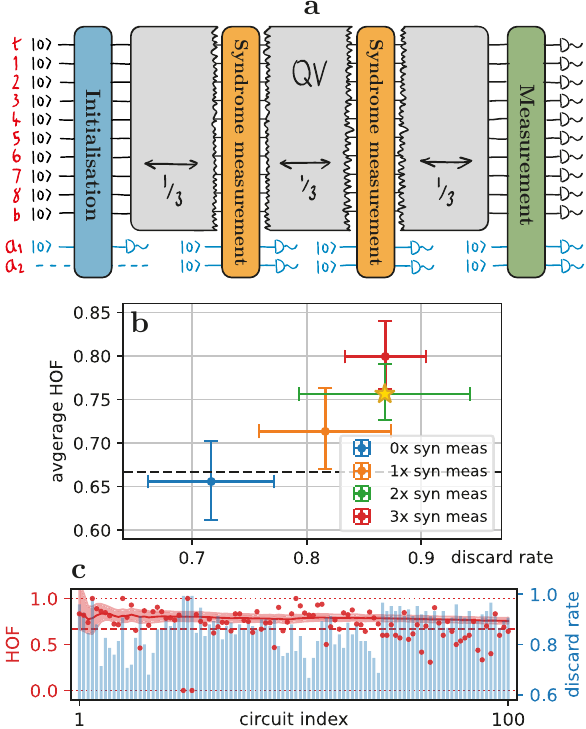}
    \caption{
    \textbf{Passing a Quantum Volume test with eight logical qubits using the Iceberg code.}
    \textbf{a}~The 8-qubit logical circuit for the quantum volume test is encoded into 10 physical qubits plus two ancillas ($a_1$ and $a_2$). The circuit includes the initialisation of all logical qubits in the $\ket{\overline{0}}$ state, a varying number of mid-circuit syndrome measurements, a final partial syndrome measurement, and the measurement of all qubits. The case of 2x mid-circuit syndrome measurements is sketched.
    \textbf{b}~Average heavy output frequency plotted against discard rate. The passing threshold 2/3 is indicated with a dashed black horizontal line. Vertical errors bars show the bounds on the mean estimate of heavy output frequency. On the horizontal axes we plot the mean discard rate with error bars showing the sample standard deviation. We run 2x syndrome measurements, indicated with a star, for the full 100 circuits required to pass the QV test. For the other QV tests we run 45 circuits instead.
    \textbf{c}~Results for the individual circuits are shown for 2x syndrome measurements. On the left axes (drawn in red) we plot the heavy output frequency of each circuit after discard, as well as the cumulative mean (increasing upwards) and its bounds. The passing threshold 2/3 is indicated with a dashed line and right axis (plotted as blue bars) shows the discard rate.
    }
    \label{fig:qvt8}
\end{figure}

In the second experiment, illustrated in Fig.~\ref{fig:qvt8}, we demonstrate a logical QV of $2^8$ and show the positive effect of increasing the number of syndrome measurements.

The QV test~\cite{cross2019validating,baldwin2022re} is a commonly used holistic benchmark of a quantum processor. It utilises random circuits with an equal number of qubits and layers. Each layer randomly pairs the qubits and applies a general $SU(4)$ gate on each pair. Each circuit is measured in its computational basis and a statistic called heavy output frequency (HOF) is computed from the results. The test is passed if the lower uncertainty bound of the average HOF is greater than 2/3, when averaged over many random circuits. We compute the bounds on HOF using a bootstrapped approach~\cite{baldwin2022re}. 

The QV circuits we use are generated using Qiskit~\cite{qiskit-pypi} and then compiled into the universal gate set of the Iceberg code. Figure~\ref{fig:qvt8}(a) shows a schematic of the circuits we run. Circuits are broken to insert a varying number of intermediate syndrome measurements. For 0x, 1x and 3x rounds of syndrome measurement we run an abridged QV test that stops at 45 circuits, where the mean HOF seem to have stabilised. The 2x case is continued to the full 100 circuits to demonstrate the passing of the QV test. The HOF for the individual circuits, as well as its cumulative mean and bootstrapped bounds are plotted for the 2x case in Fig.~\ref{fig:qvt8}(c).

As shown in Fig.~\ref{fig:qvt8}(b) adding syndrome measurements increases the HOF and consequently increases the confidence in passing the QV test. This improvement comes at the cost of an increased discard rate. As expected, the increase in the discard rate moderates as it approaches the maximum discard rate of 100\%.

\section{Discussion}
Early fault-tolerant quantum computation (EFTQC) protocols aim to incorporate cheap QEC and QED primitives that fit within the limitations of current quantum computers and provide some level of protection against noise. We contribute to EFTQC with the development of the Iceberg code, a $[[k+2, k, 2]]$ QED code that employs the minimal quantum overhead to protect arbitrarily large expressive quantum circuits. Due to its structure, the Iceberg code is ideally suited to hardware with all-to-all connectivity and arbitrary two-qubit rotations such as trapped-ion devices.

With experiments on the Quantinuum H1-2 trapped-ion quantum computer, we demonstrate the Iceberg code's ability to improve the quality of results from expressive quantum circuits. In our experiments the code protects parameterised quantum mirror circuits varying in depth from 4 to 256 layers. Further, we achieve a logical quantum volume of $2^8$ with a discard rate of $86.8 \pm 0.7\%$. This saturates the logical QV that can be achieved in this device with the Iceberg code. Remarkably this demonstration requires circuits with up to 379 two-qubit gates when compiled into native hardware operations.
Finally, we show that adding syndrome measurements interspersed throughout the circuits further increases the quality of the results, an important feature of QEC/QED.

As expected in QED, we observe that increasing the number of logical qubits, the circuit depth, or the level of protection via more syndrome measurements comes at the cost of an increasing discard rate for the Iceberg code. The price to pay is then an overhead in the number of circuit repetitions. In our experiments all circuits run to the end and error detection is implemented as a post-selection. However, it is possible to implement a conditional exit during runtime as soon as an error is detected. We expect this to halve the experimental runtime of a rejected circuit on average.

There are many other aspects of the code that are yet to be explored. Optimised hardware and compilation strategies~\cite{Itoko_2020, Gokhale_2020, Lao_2021, Nannicini_2021} can boost the level of protection that this code offers against noise and reduce the discard overhead. In the same spirit of improving the level of protection, the Iceberg code can correct for one qubit loss using existing protocols demonstrated on trapped-ion devices~\cite{Stricker_2020}. Coherent errors are another potential noise source that should studied in depth. Unlike the QV experiment, the structure of our mirror circuits potentially makes them insensitive to coherent errors. Hence, more sophisticated benchmarking and additional mitigation techniques can help to improve the protection that the Iceberg code offers against this particular noise.

In the context of applications, the code offers a particularly high-fidelity compilation of highly expressive parameterised quantum circuits of the form used by variational quantum algorithms in optimisation~\cite{Farhi_2014, Amaro_2022} and machine learning~\cite{Benedetti_2019}. In particular, the presence of high-fidelity global logical gates can open the door to shallower compilations~\cite{Wetering_2021} and longer time Hamiltonian simulation by Trotterisation.

We conclude that the Iceberg code and its potential applications fit centrally within the EFTQC paradigm by explicitly embracing the trade-off between usability and fault-tolerance. 

\section{Acknowledgments}
We thank Ciar\'{a}n Ryan-Anderson, Nathaniel Burdick, Brian Neyenhuis, Charlie Baldwin, Yuta Kikuchi, Mattia Fiorentini, David Hayes, Karl Mayer and Yi Hsiang Chen for fruitful discussions and feedback on the manuscript. The experiments were done using the Quantinuum system model H1-2, powered by Honeywell ion traps.

\section{Author contributions}
All authors conceived and designed the study. D.A. performed analytic calculations, C.S. carried out numerical studies. All authors analysed the data, interpreted the results and wrote the manuscript.

\section{Data availability}
The data that support the findings of this study are available at Zenodo~\cite{self_chris_2023_8318683}.

\bibliography{main}
\bibliographystyle{naturemag}

\clearpage

\onecolumngrid

\begin{center}
\large
\noindent\textbf{Supplementary Information for\\``Protecting Expressive Circuits with a Quantum Error Detection Code''}
\end{center}

\makeatletter
\setcounter{figure}{0}
\renewcommand{\fnum@figure}{\textbf{Supplementary Figure \thefigure}}
\makeatother

\section{Iceberg codes}
Here we give additional details on the $[[k+2, k, 2]]$ code, which we have nicknamed the Iceberg code. These supplement the descriptions of the stabilisers, logical operators and universal gate set given in the main text.

\subsection{Logical operators}
\label{app:code-defns}

One of the useful properties of the Iceberg code is there are global logical operators compiled onto a corresponding local two-qubit physical operator. For simplicity let us label the qubits into the set $[k]=\{1,\ldots,k\}$ and the set $[n]=[k]\cup\{t,b\}$. Here we list all logical operators sorted by their two-qubit physical compilation:
\begin{align}
    \overline{X}_i &= X_t X_i \quad \forall \, i\in [k] \label{eqn:code-expansion-start}  \\
    \overline{X}_i \overline{X}_j &= X_i X_j \quad \forall \, i,j\in [k] \\
    \otimes_{j \in [k]\setminus i} \overline{X}_j &= X_b X_i \quad \forall \, i\in [k] \label{eqn:code-expansion-global1} \\
    \otimes_{j \in [k]} \overline{X}_j &= X_t X_b \label{eqn:code-expansion-global2}
\end{align}
\begin{align}
    \overline{Z}_i &= Z_b Z_i \quad \forall \, i\in [k] \\
    \overline{Z}_i \overline{Z}_j &= Z_i Z_j \quad \forall \, i,j\in [k] \\
    \otimes_{j \in [k]\setminus i} \overline{Z}_j &= Z_t Z_i \quad \forall \, i\in [k] \label{eqn:code-expansion-global3} \\
    \otimes_{j \in [k]} \overline{Z}_j &= Z_t Z_b \label{eqn:code-expansion-global4}
\end{align}
\begin{align}
    \overline{Y}_i \overline{Y}_j &= Y_i Y_j \quad \forall \, i,j\in [k] \\
    \overline{X}_i \otimes_{j \in [k]\setminus i} \overline{Z}_j &= - Y_t Y_i \quad \forall \, i\in [k] \label{eqn:code-expansion-global5} \\
    \overline{Z}_i \otimes_{j \in [k]\setminus i} \overline{X}_j &= - Y_b Y_i \quad \forall \, i\in [k] \label{eqn:code-expansion-global6} \\
    \otimes_{j \in [k]} \overline{Y}_j &= (-1)^{1+k/2} Y_t Y_b. \label{eqn:code-expansion-end}
\end{align}
We see that both single- and two-qubit logical operators are compiled into a two-qubit physical operator. Besides, in Eqs.~\eqref{eqn:code-expansion-global1},~\eqref{eqn:code-expansion-global2},~\eqref{eqn:code-expansion-global3},~\eqref{eqn:code-expansion-global4} and~\eqref{eqn:code-expansion-global5}-\eqref{eqn:code-expansion-end} we see that global logical gates acting on $k$ or $k-1$ logical qubits are also compiled into a two-qubit physical operator. 

\subsection{Fault-tolerant operations}
\label{app:fault-tolerance}

The Iceberg code has fault-tolerant operations for initialisation and syndrome measurement. These circuits require two additional ancilla qubits, which can be reset and reused. Our physical circuits therefore require $k+4$ physical qubits, giving a constant overhead.

\begin{figure*}
    \centering
    \includegraphics[width=0.89\columnwidth]{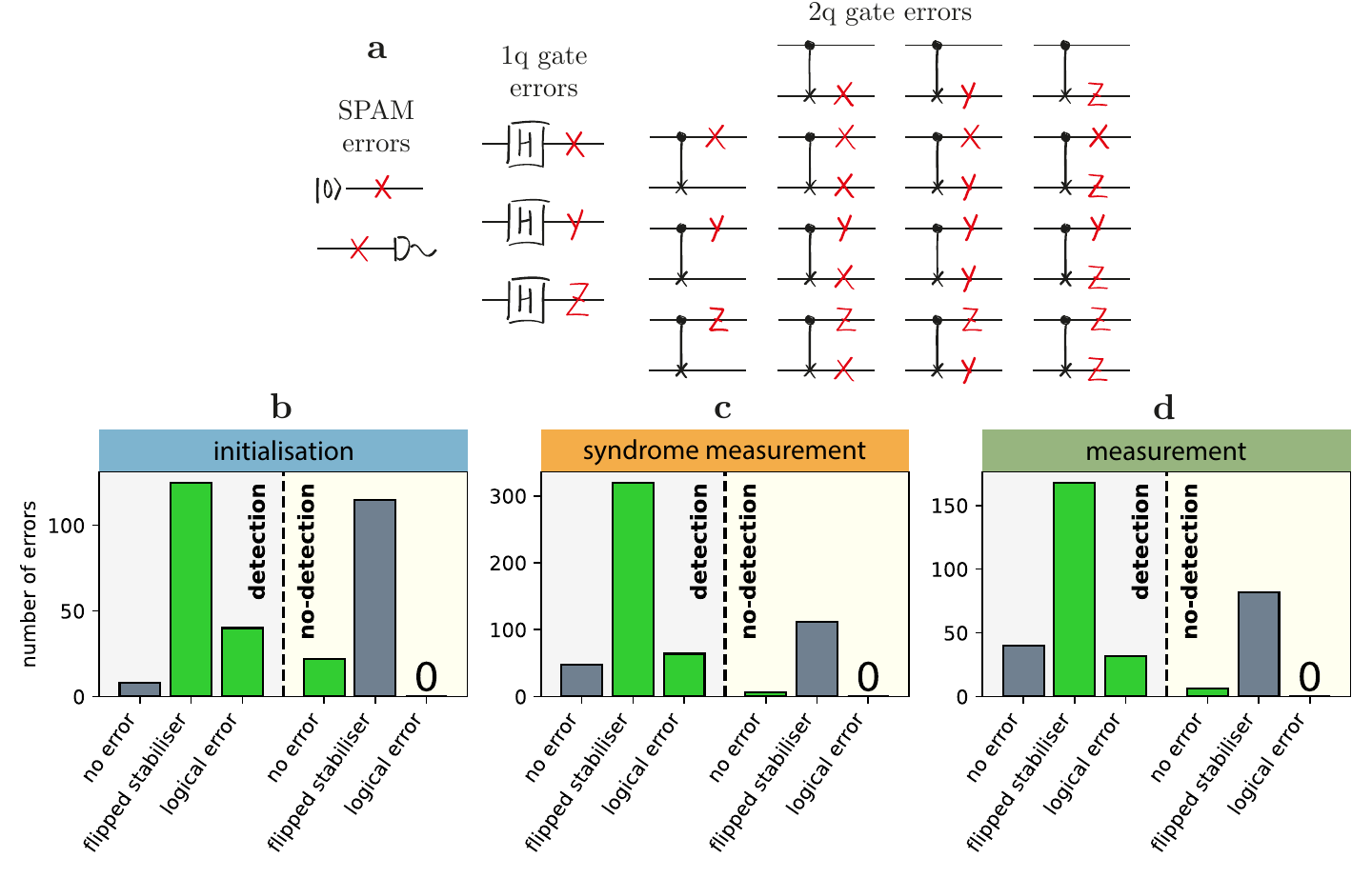}
    \vspace{-12px}
    \caption{
    Verification that the initialisation, syndrome measurement and measurement circuits of the Iceberg code are fault-tolerant for $k=16$ logical qubits. 
    \textbf{a} The full set of local errors that the initialisation, syndrome measurement and measurement circuits (shown in Fig.~\ref{fig:ft-circs} of the main text) are fault-tolerant against.
    \textbf{b},\textbf{c},\textbf{d} For each circuit, all local errors are classified based on whether they trigger a detection readout (left three bars), or do not (right three bars). Within these groups the error can correspond to no error (for example an error with the form of a stabiliser), a flipped stabiliser, or a logical error. We colour the cases based on whether they are ideal behaviour (green) or non-ideal behaviour that is not a fail (grey). The three circuits are fault-tolerant because the number of undetected logical errors is zero.}
    \label{fig:ft-verification}
\end{figure*}

The circuits we implement for these tasks are fault-tolerant in the sense that there are no single faults that give rise to an undetectable logical error. We define single faults to be local Pauli errors arising from the failure of a circuit element. The full set of local errors we consider is shown in Supp. Fig.~\ref{fig:ft-verification}(a).

Initialisation prepares the \emph{all-zeros} code state, which is the state such that $\overline{Z}_i \ket{\overline{0}}^{\otimes [k]} = +1 \ket{\overline{0}}^{\otimes [k]}$ for all $i \in[k]$. 
From the definition of the stabilisers and logical operators we see that this state is the Greenberger–Horne–Zeilinger (GHZ) state of the $n$ physical qubits:
\begin{equation*}
    \ket{\overline{0}}^{\otimes [k]} = \frac{1}{\sqrt{2}} \left(\ket{0}^{\otimes [n]} + \ket{1}^{\otimes [n]} \right)  .
\end{equation*}
The fault-tolerant circuit we use to create this state is shown in the main text in Fig.~\ref{fig:ft-circs}(c). 

The syndrome measurement should enable readout of the stabiliser eigenvalue without disturbing the logical information. The syndrome measurement circuit we use is drawn in the main text in Fig.~\ref{fig:ft-circs}(d), where two ancillas are used to separately readout the two stabilisers. The operations are arranged so that the ancillas act as flag qubits to each other.

We perform a destructive measurement at the end of the circuit by measuring all physical qubits in their computational basis. This allows us to reconstruct the measurement outcome for the $\overline{Z}_i$ logical operators and the $S_Z$ stabiliser. However, we cannot reconstruct the $S_X$ stabiliser from these measurements. To give us this information we extract the $S_X$ stabiliser before the final readout. This is done using a flagged fault-tolerant circuit~\cite{Chao_2008}, shown in the main text in Fig.~\ref{fig:ft-circs}(e).

We verify that {initialisation}, {syndrome measurement} and measurement circuits are fault-tolerant by exhaustively checking all code sizes up to $k=16$. For each possible local error, we insert the corresponding Pauli operator, propagate it to the end of the circuit, and finally assess its effect. We confirm that there are no errors that give rise to an undetectable logical error. This is illustrated for the case of $k=16$ in Supp. Fig.~\ref{fig:ft-verification}(b-d), which shows a classification of all local errors. 

\subsection{Compiling target unitaries into a logical circuit}

A generic approach to compiling a target unitary $\overline{U}$ into a logical circuit is to first compile $\overline{U}$ into a gate set that only contains single-qubit rotations, $\overline{R}_{\overline{X}}(\theta) = \exp(-i \theta \overline{X} / 2)$ and $\overline{R}_{\overline{Z}}(\theta) = \exp(-i \theta \overline{Z} / 2)$, and two-qubit logical rotations, $\overline{R}_{\overline{X} \overline{X}}(\theta) = \exp(-i \theta \overline{X} \overline{X} / 2)$, $\overline{R}_{\overline{Y} \overline{Y}}(\theta) = \exp(-i \theta \overline{Y} \overline{Y} / 2)$ and $\overline{R}_{\overline{Z} \overline{Z}}(\theta) = \exp(-i \theta \overline{Z} \overline{Z} / 2)$.
This can be done using the TKET compiler~\cite{sivarajah2020t,pytket-pypi}. Once expressed using single and two-qubit logical gates, each gate in the circuit can be directly compiled into its encoded physical form using $\text{MS}_{ij}(\theta)$ gates along with up to four single-qubit Clifford gates.

\section{Mirror Circuit Experiments}

Our mirror circuit experiments use 8 logical qubits encoded into 10 physical qubits. Experiments were executed on the Quantinuum H1-2 trapped-ion quantum computer between the 13th September and 4th October 2022.

In the following we discuss the experiments in more detail. First, we give a full discussion of the construction of the random mirror circuits, followed by additional practical details of the experiment. Next we give an in-depth account of the numerical simulations presented in the main text including the simplified noise model employed and the simulation method. Finally, we present numerical evidence that the random circuits used in our experiments are typical of the random circuit families we construct.

\subsection{Random circuit construction}

\begin{figure*}
    \centering
    \includegraphics[width=0.92\columnwidth]{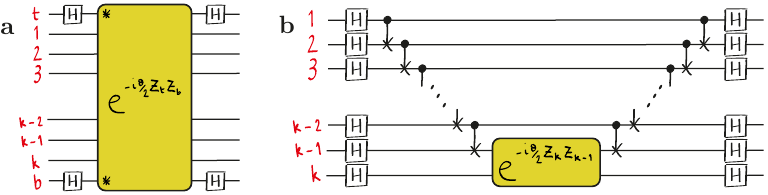}
    \caption{
    Compilations of the rotation $R_{X t b}(\theta) = \exp(-i \theta \overline{P}_{X t b}/2)$, generated by $\overline{P}_{X t b} = \otimes_{i \in [k]} \overline{X}_i = X_t X_b$. 
    \textbf{a} With the Iceberg code, the rotation is realised by a single two-qubit $\text{MS}_{t b}(\theta)$ gate, as well as four Hadamard gates.
    \textbf{b} In the unencoded circuit, the physical circuit for the same rotation utilises a phase gadget. Since the logical generator of the rotation is a global operator, $\otimes_{i \in [k]} \overline{X}_i$, a ladder of CNOT gates is used to compute the parity before applying a single $\text{MS}_{k(k-1)}(\theta)$ gate. Then another CNOT ladder is used to uncompute the parity. Additionally, Hadamard gates on all qubits at the start and end apply a basis change.
    }
    \label{fig:mirror-circs-app}
\end{figure*}

As described in the main text, random unitaries are constructed starting from the encoded physical circuit. 
The unitaries are built in layers. Within each layer physical qubits are paired $(i, j)$ for $i \neq j \in [ n ]$ to give $\tfrac{k}{2} + 1$ pairings. For each pairing we uniformly choose $\sigma \in \{X, Y, Z\}$, uniformly sample an angle $\theta \in (0, 2\pi)$, and apply a two-qubit rotation to the pair, of the form:
\begin{equation}
    R_{\sigma i j}(\theta) = \exp(-i \theta \sigma_i \sigma_j /2)  .
\end{equation} 
Logical gates are all physically compiled using a $\text{MS}_{ij}(\theta) = \exp(-i \theta Z_i Z_j /2)$ gate along with up to four single-qubit Clifford gates. An example of the physical compilation of $R_{X t b}(\theta)$ is shown in Supp. Fig.~\ref{fig:mirror-circs-app}(a).

Once the physical random unitary is constructed we map it to the corresponding logical unitary.
All physical operators $\overline{P}_{\sigma i j} = {\sigma}_i {\sigma}_j$ represent a logical Pauli operator, the full set of relations is given in Eqs.~\eqref{eqn:code-expansion-start}-\eqref{eqn:code-expansion-end}. Logical operators can be single- and two-qubit rotations or rotations with generators that act on $k$ or $k-1$ qubits. Single-qubit rotations are compiled directly. Two-qubit rotations are compiled using the two-qubit rotation $\text{MS}_{ij}(\theta)$ along with up to four single-qubit Clifford gates. Global rotations on $k$ or $k-1$ qubits are compiled with a phase gadget whose kernel is a $\text{MS}_{ij}(\theta)$ gate. Supplementary Figure~\ref{fig:mirror-circs-app}(b) shows an example of the compiled $R_{\otimes_{i\in[k]}\overline{X}_i}(\theta)$ in the unencoded circuit.

\subsection{Additional experimental details}

Circuits are initially constructed in the gateset: $\{ H, S, S^\dagger, \textrm{CNOT}, \textrm{MS}(\theta) \}$, using the construction of the initialisation, syndrome measurement and final measurement circuits given in Fig.~\ref{fig:ft-circs} of the main text as well as the construction of the random circuits described previously. These circuits are then compiled into the native gateset of the Quantinuum hardware~\cite{Pino_2021} using the TKET compiler~\cite{pytket-pypi}.

We consider from 4 random layers up to a maximum of 256. When reporting the number of layers we count the layers in both the unitary and its inverse, in order to count the total number of layers between initialisation and final measurement. In our experiments on circuits encoded with the Iceberg code we insert one round of syndrome measurement between the unitary and its inverse for 32, 64, 128 \& 256 layers, while we perform no intermediate syndrome measurements for 4, 8 \& 16 layers.

Circuits are run for different numbers of repeats, here we report the total number of repeats (measurment shots) run in each case. For unencoded circuits with global rotations all cases have 300 shots; without global rotations 4, 8 \& 16 layers run with 300 shots, 32 layers with 400, 64 layers with 600, 128 layers with 800 and 256 layers with 500. For the encoded circuits both with and without global rotations use the same shot numbers: 4, 8, 16 \& 32 layers run with 200 shots, 64 layers with 400, 128 layers with 800 and 256 layers with 500. The approximate total experimental runtimes, including both encoded and unencoded circuits, is 7 hours for circuits containing global rotations and 8 hours for circuits not containing global rotations.

\subsection{Numerical simulations}

The numerical simulations in the main text cover $k=10, 12, 14, 16$ logical qubits. For each $k$, we consider increasingly deep circuits, from $l=4$ layers up to a maximum of 128. At each $l$, we generate 32 random mirror circuits allowing global logical rotations. Syndrome measurement rounds are inserted after every 16 layers. The noisy output of these circuits is simulated using a Monte-Carlo statevector approach described below. For each circuit we draw 32 statevector samples from the error model. We simulate both the unencoded and encoded version of each random circuit and compute the survival probabilities exactly for each statevector. Averaging the survivial probabilities over these 32 statevector samples gives the estimated survivial probability of the random circuit.

We simulate a simplified noise model that includes state preparation and measurement (SPAM) errors, as well as depolarising errors acting after gates. SPAM errors are represented with a bit-flip channel applied after qubit initialisation and before measurements, 
\begin{equation*}
    \mathcal{E}_\textrm{bitflip}(\rho) = (1 - p_b) \rho + p_b \, X \rho X \, .
\end{equation*}
After each single (1q) and two-qubit (2q) gate we apply a depolarising error channel to the qubits acted on by the gate. The operator-sum expressions for these depolarising channels are
\begin{equation*}
    \mathcal{E}_\textrm{1q}(\rho) = (1 - p_{1q}) \rho + \frac{p_{1q}}{3} \left(  X \rho X + Y \rho Y + Z \rho Z \right)
\end{equation*}
\vspace{-12pt}
\begin{equation*}
\begin{split}
    \mathcal{E}_\textrm{2q}(\rho) = (1 - p_{2q}) \rho + \frac{p_{2q}}{15} & \bigg( I X \rho I X + I Y \rho I Y + I Z \rho I Z \\
    & + X I \rho X I + X X \rho X X + X Y \rho X Y + X Z \rho X Z \bigg. \\
    & + Y I \rho Y I + Y X \rho Y X + Y Y \rho Y Y + Y Z \rho Y Z \bigg. \\
    & + Z I \rho Z I + Y X \rho Z X + Z Y \rho Z Y + Z Z \rho Z Z \bigg)
\end{split}
\end{equation*}
where $I$ is the identity and $X$, $Y$, $Z$ are the Pauli matrices. 

Based on randomised benchmarking of the Quantinuum H1-2 hardware we use the following values for the parameters of the error channels: initialisation errors have a bitflip error rate $p_b = 4\times10^{-4}$, measurement errors have a bitflip error rate $p_b = 3\times10^{-3}$, single qubit gate errors have a depolarising rate $p_{1q} = 4\times10^{-4}$ and two qubit gate errors have a depolarising rate $p_{2q} = 3\times10^{-3}$.

Our noisy simulations are Monte-Carlo statevector simulations. For each noise channel we randomly select one of the Kraus operators of the channel with the correct probabilities. The noise channel is replaced with this randomly chosen Kraus operator. Once this is done for all error channels this turns the noisy quantum evolution back into a unitary operator that can be studied with a statevector simulator. By repeatedly sampling different Kraus operators of the noise channels we can approximate the full noisy evolution. This makes larger system sizes accessible by avoiding working with the, exponentially larger, density matrices.

In our quantum error detection circuits we carry out mid-circuit measurements and post-selection. These are implemented in our state-vector simulations with projectors in the computational basis. This allows us to compute both the probability of discard at that mid-circuit measurement and the normalised post-selected statevector (assuming the probability of discard is less than one).

The numerical results presented in the main text are obtained using a Qiskit~\cite{qiskit-pypi} statevector simulator. 
The circuits are prepared in the gateset: $\{ H, S, S^\dagger, \textrm{CNOT}, \textrm{MS}(\theta) \}$, following the construction of the initialisation, syndrome measurement and final measurement circuits given in Fig.~\ref{fig:ft-circs} of the main text along with the random circuits described previously. We consider the noise model acting on the circuits in this gateset.

In both the unencoded and encoded cases the survival probability is computed exactly from each statevector. In the unencoded case, this is simply the probability amplitude of the all-zeros computational basis state. In the encoded case, we must first extract the full probability distribution of outcomes in the computational basis, we then decode each computational basis state and add up the contributions to the logical all-zeros state.

\subsection{Typicality of the instances used in experiments}

In our experimental results we consider single instances of the random circuit families we describe. For a fixed number of layers, we generate one random circuit for the `with global rotations' case and another random circuit for the `without global rotations' version. The circuits generated at different numbers of layers are independent of each other. Here we give numerical evidence that these instances are typical instances of the random circuit families. 

\begin{figure*}
    \centering
    \includegraphics[width=0.86\columnwidth]{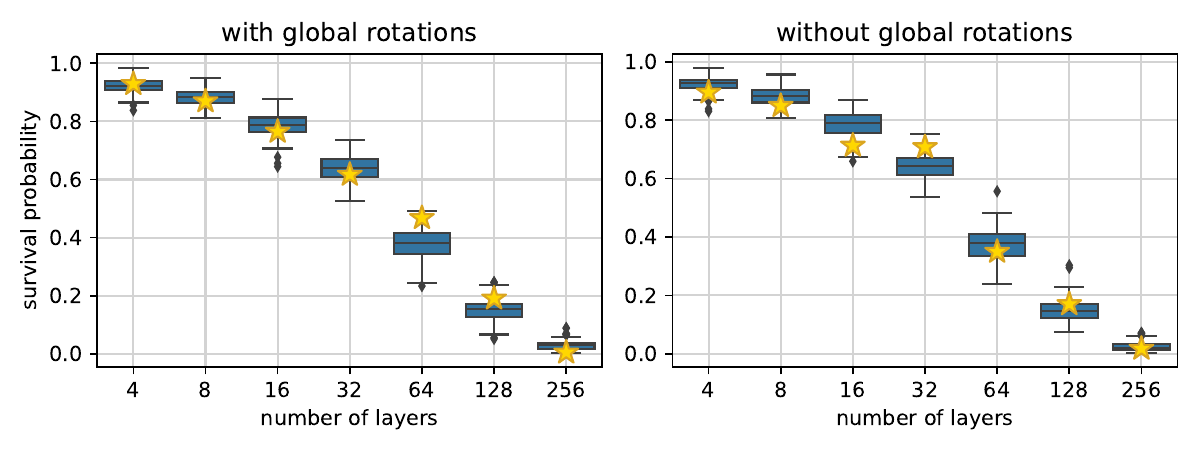}
    \vspace{-8px}
    \caption{
    Distribution of the survival probabilities at each circuit depth for 100 random instances (left) including global rotations and (right) without global rotations, using box plots.
    The centre line of the box plots shows the median, with the lower and upper box edges corresponding to the 1st and 3rd quartile respectively. The whiskers of the plot extend to show data up to 1.5 $\times$ the interquartile range below the 1st quartile and above the 3rd quartile. Outliers beyond these limits are plotted individually.
    Star icons indicate the performance of the experiments we present in the main text.
    Results are obtained from noisy simulations using a simplified noise model.
    }
    \label{fig:typicality-data}
\end{figure*}

At each depth we generate 100 instances of the two random circuit families and Supp. Fig.~\ref{fig:typicality-data} plots the distribution of the survival probabilities we observe as a box plot. Outliers are identified as points that lie more than 1.5 times the interquartile range above(below) the upper(lower) quartile.
Each instance is numerically simulated using the simplified error model and statevector simulation method described previously under ``Numerical Simulations''. For each circuit we average over 100 statevector samples from the error model.

The random instances that we use in experiments are indicated in Supp. Fig.~\ref{fig:typicality-data} with star icons. These cases correspond to fixed seeds for random generators. We see that none of these instances correspond to outliers and they are spread above and below the median.

\section{Logical Quantum Volume Experiments}

We have carried out a logical Quantum Volume (QV) test of 8 qubits encoded in 10 physical qubits using the Iceberg code running on Quantinuum's H1-2 trapped-ion quantum computer. The data presented in the main text was obtained between the 27th July and the 8th October 2022. 

Here we give more details of these experiments -- beginning with a short review of the QV test, followed by a discussion of the preparation of the logical circuits including the optimisations we apply. Next we discuss the encoding of the QV circuits into the Iceberg code. Finally, we give further practical details of the experiment and present additional experimental results.

\begin{figure*}
    \centering
    \includegraphics[width=0.72\columnwidth]{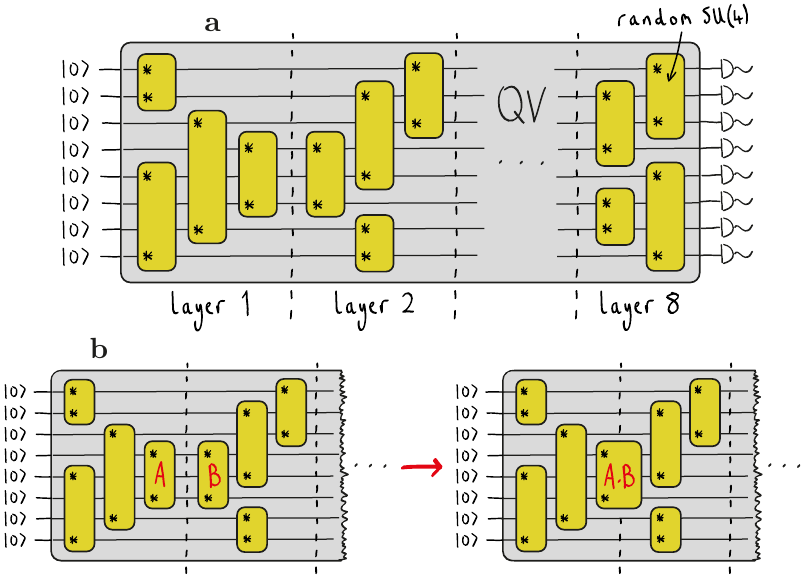}
    \caption{\textbf{a} Schematic example of a Quantum Volume (QV) circuit for $N=8$ qubits. The circuit consists of $N$ layers, within each layer the qubits are randomly paired and the pairs are acted on by a random $SU(4)$. \textbf{b} Compression of $SU(4)$ blocks in QV circuits. We show a schematic of the first two layers of a six qubit QV test, where $A$ and $B$ are random $SU(4)$ unitaries. Qubits 4 and 6 are randomly paired in both layers meaning $A$ and $B$ act on the same qubits. They can be combined into a single $SU(4)$ unitary. This will generally result in a shorter circuit depth than applying $A$ and $B$ separately.}
    \label{fig:qv-app-circuits}
\end{figure*}

\subsection{QV test summary}
\label{app:qv-summary}

The QV test is a sample efficient, randomised sampling scheme to verify that non-classical processing is happening inside a noisy quantum computer~\cite{cross2019validating,baldwin2022re}.
The test applies random square circuits of the form shown in Supp. Fig.~\ref{fig:qv-app-circuits}(a) to $k$ qubits. These circuit have $k$ layers, within each layer the qubits are randomly paired and a random $SU(4)$ unitary is applied to that pair. Finally, all of the qubits are measured in the computational basis.

Each random circuit is simulated noiselessly to obtain the ideal output distribution and from the ideal distribution we determine the set of heavy outputs. These are computational basis states that occur with higher probability than the median. 
The circuit is executed on a noisy quantum processor with a small number of shots (e.g. 100). The heavy output frequency of the noisy output is computed by counting up the number of shots that measure a computational basis state from the set of heavy outputs.

Heavy output frequency is averaged over the random circuits and the QV test is passed if the average is $> 2/3$. If a device passes a QV test of $k$ qubits, it has quantum volume $2^k$.
Average heavy output frequency is not a fidelity measure, for noiseless circuits it is $\approx 85$\% and for completely depolarised states it is 50\%.

\subsection{Preparation of the logical QV circuits}
\label{app:qv-prep-logical}

We use quantum volume circuits generated from Qiskit~\cite{qiskit-pypi} and implement the `medium' optimisation method described in Ref.~\cite{baldwin2022re}. Using this method $SU(4)$ blocks are combined together when allowed. An example of when this would occur is shown in Supp. Fig.~\ref{fig:qv-app-circuits}(b).

Following these optimisations, the $SU(4)$ blocks are decomposed into single-qubit rotations -- $R_{\overline{X}}(\theta)$ and $R_{\overline{Z}}(\theta)$ -- and two-qubit logical rotations -- $R_{\overline{X} \overline{X}}(\theta)$, $R_{\overline{Y} \overline{Y}}(\theta) $, $R_{\overline{Z} \overline{Z}}(\theta)$.
This is done using the TKET compiler~\cite{sivarajah2020t,pytket-pypi}. 
At this stage we further optimise by squashing together single-qubit logical rotations as much as possible. For example, a sequence of gates $R_{\overline{Z}}(a) R_{\overline{X}}(b) R_{\overline{Z}}(c) R_{\overline{Z}}(d) R_{\overline{X}}(e) R_{\overline{Z}}(f)$ can be replaced with a shorter sequence $R_{\overline{Z}}(\alpha) R_{\overline{X}}(\beta) R_{\overline{Z}}(\gamma)$ that applies the same logical rotation. Additionally, we can remove Pauli-$Z$ rotations that act immediately after initialisation and immediately before measurement.

\subsection{Encoding the QV circuits}
\label{app:qv-encoding}

Once expressed using single and two-qubit logical rotations the QV circuit can be directly translated into its physical form, using Eqs.~\eqref{eqn:code-expansion-start}-\eqref{eqn:code-expansion-end}.
This logical circuit is then combined with initialisation and measurement instruction sets. We note that due to the structure of the Iceberg code, where all single-qubit logical $X$ rotations involve the physical qubit $t$ and all single-qubit logical $Z$ rotations involve the physical $b$, the compiled physical circuit does no longer present the parallel form of the logical QV circuit.

Syndrome measurements are inserted between equally sized chunks of the QV instruction set. When doing this we do not respect the layered structure of the QV circuit, meaning the syndrome measurement circuits may be inserted inside a QV layer.
However, we must be careful not to break up logical operations, e.g. to insert a syndrome measurement between the Hadamards and $\text{MS}_{ij}(\theta)$ gate of $\exp(-i X_i X_j /2) = (H_i H_j)  \text{MS}_{ij}(\theta)  (H_i H_j)$.

The final executable circuits for the $k=8$ logical QV test act on 12 physical qubits and have circuit depths of $\sim700$ up to over 1000 depending on the number of syndrome measurements, including up to $\sim350$ two-qubit gates. 

\begin{figure*}
    \centering
    \includegraphics[width=\columnwidth]{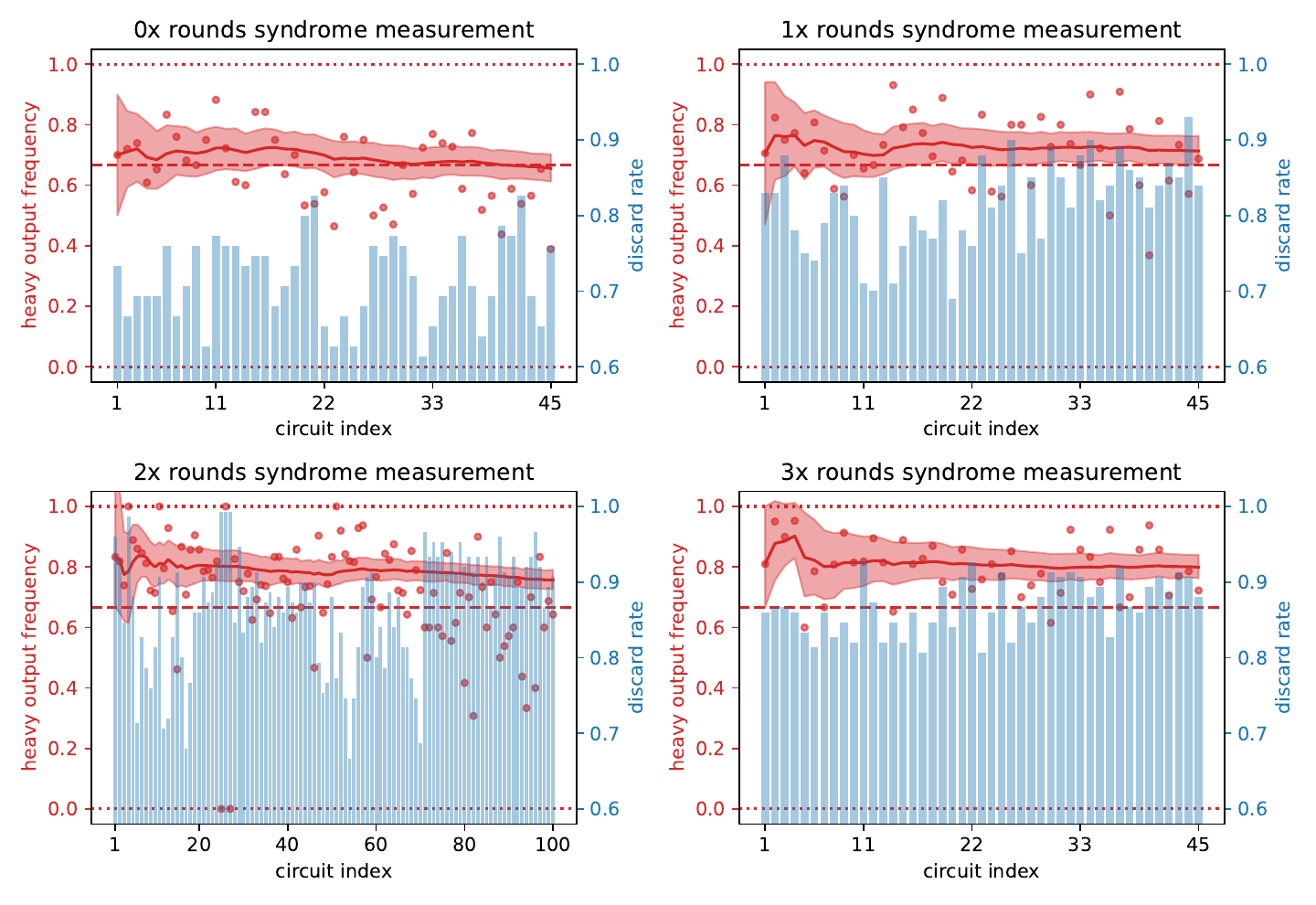}
    \vspace{-12px}
    \caption{
    Additional experimental results showing the heavy output frequency of individual Quantum Volume (QV) circuits for 0x, 1x, 2x and 3x syndrome measurements. In each panel, the left axes (drawn in red) plots the heavy output frequency of each circuit after discarding, as well as the cumulative mean and its bootstrapped bounds. The passing threshold 2/3 is indicated with a dashed line, as well as 0 and 1 being highlighted with dotted lines. The right axes (plotted as blue bars) shows the number of shots retained after discarding. Experiments with 0x syndrome measurements used 75 shots, with 1x syndrome measurement used 100 shots, and with 2x and 3x syndrome measurements used 150 measurement shots.
    }
    \label{fig:qv-multi-series}
\end{figure*}

\subsection{Additional experimental details and results}
\label{app:qv-experimental-details}

Circuits are initially constructed in the gateset: $\{ H, S, S^\dagger, \textrm{CNOT}, \textrm{MS}(\theta) \}$, following the construction of the initialisation, syndrome measurement and final measurement circuits given in Fig.~\ref{fig:ft-circs} of the main text as well as the construction of the encoded QV circuits described previously. These circuits are then compiled into the native gateset of the Quantinuum hardware~\cite{Pino_2021} using the TKET compiler~\cite{pytket-pypi}. 

Our experiments use 0x, 1x, 2x, 3x syndrome measurements. For 0x rounds of syndrome measurement we carry out 75 repeats of the experiment (measurement shots), for 1x we use 100 shots and for 2x and 3x we perform 150 shots. We vary the number of shots with the goal of retaining 10-20 shots of each circuit after discards, while the discard rate varies with the number of syndrome measurements. For the 0x, 1x and 3x cases we run 45 random circuits, while for 2x we run 100 circuits. The approximate total experimental runtimes is 5 hours for 0x rounds of syndrome measurement, 6 hours for 1x rounds, 20 hours for 2x rounds and 11 hours for 3x rounds.

All circuits run to the end, with the outcome of mid-circuit error detection measurements being saved to classical syndrome registers. When the measurement results are collected we discard results where an error is detected. The results that are not discarded are post-processed to reconstruct the values of the logical operators $\overline{Z}_i$ as well as $S_Z$. If $S_Z$ detects an error we discard the result. The remaining decoded results are processed to compute the heavy output frequency.

The heavy output frequencies for each individual QV circuit with 0x, 1x, 2x and 3x syndrome measurements are shown in Supp. Fig.~\ref{fig:qv-multi-series}, as well as the discard rates.
The reported cumulative averages are calculated using Qiskit~\cite{qiskit-pypi}. The bounds shown are calculated using the bootstrapped approach described in Ref.~\cite{baldwin2022re}.

\end{document}